\documentstyle[preprint,aps] {revtex}
\begin{document}
\draft

\title {Thermal conductivity and gap structure of the 
superconducting phases of UPt$_3$}
\author{H. Suderow, J.P. 
Brison*, A. Huxley and J. Flouquet}
\address{D\'{e}partement de 
Recherche Fondamentale sur la 
Mati\`{e}re \\ Condens\'{e}e, SPSMS,CEA/Grenoble, 17 rue 
des 
Martyrs, 38054 
Grenoble cedex 9, France.}
\address{* Centre de Recherches sur les Tr\`{e}s Basses 
Temp\'{e}ratures, CNRS, BP 166,
38042 Grenoble-Cedex 9, France}
\date{\today}

\maketitle

\begin{abstract}
We present new measurements of the thermal conductivity 
($\kappa$) of UPt$_3$ down to very low temperatures (16mK) and 
under magnetic fields (up to 4 T) which cover all the 
superconducting phases of UPt$_3$. The measurements in zero field 
are compared with recent theoretical predictions for the thermal 
conductivity, which is dominated by impurity states at the lowest 
temperatures studied. The measurements under magnetic field at 
low temperatures are surprising since they don't show the expected 
low field square root dependence, $\kappa\propto\sqrt{B}$. The 
discontinuity of $d\kappa/dT$ at T$_c$ changes drastically when 
passing from the high field low temperature C phase to the low field 
high temperature A phase : this is related to the change of the 
symmetry of the superconducting order parameter when crossing 
the A$\rightarrow$ C phase transition.

\
\
\
\
\center{To be published in Journal of Low Temperature Physics}

\end{abstract}

\vspace{0.5cm}


\section{Introduction}
UPt$_3$ is widely believed to be an unconventional superconductor. 
The  existence of several different superconducting phases in the B-
P-T phase  diagramm is experimentally well confirmed, and can 
only be explained by a change of the symmetry of the order  
parameter (OP) when passing from one superconducting phase to 
another (for reviews, see e.g. Ref.\cite{Reviews}). This, in turn, can 
only be understood if additional symmetries, other than gauge 
symmetry, are broken at the superconducting transition. The 
hexagonal crystal structure of this compound  restricts
 the possible OP to 8 one-dimensional (so 
called A,B) and
4	 two-dimensional (E) irreducible representations of the crystal 
point group, which are
further classified  according to the parity of the superconducting 
wave functions (index u for odd and g for even parity) 
\cite{Reviews}.
 In some of the irreducible representations the order parameter 
changes sign under some of the point group symmetries, implying 
points or lines of zeros of the gap on the Fermi surface. Much 
theoretical and experimental work has been  done in order to try to 
find the symmetry of the OP in this compound, but up to  now,
 no definite consensus emerges.
The thermal conductivity is sensitive to the  position and type of 
the nodes
(line or point nodes) of the  superconducting gap, which itself 
depends on the symmetry of the OP.
Previous work (See Refs. 
\cite{Sulpice86,Behnia91,Huxley95,Suderow95,Lussier94,Lussier95}
for experimental and Refs. 
\cite{Pethick86,Schmitt86,Monien87,Arfi88,Hirschfeld86,Hirschfeld88,Graf96,Graf96_2,Fledderjohann95,Norman96,Barash96} for 
theoretical work) has already demonstrated that the thermal 
conductivity ($\kappa$) in this
compound is of electronic  origin at very low temperature, and thus 
sensitive to the
electronic  excitations in the superconducting state. Compared to the 
specific heat C,  $\kappa$
is a directional probe, and it seems \cite{Huxley95,Suderow95} that 
the  anomaly present in the specific heat below 100mK does not 
influence  $\kappa$. Thus, a directional study of $\kappa$ down to 
very low temperatures is expected  to give much information about 
the gap structure of UPt$_3$. Note that the first measurements 
\cite{Behnia91,Huxley95,Suderow95,Lussier94,Lussier95} had 
already stressed the necessity
of choosing a model for the OP yielding a hybrid gap with a line node in 
the basal plane and point node along the c axis. Measurements on 
far improved samples to lower temperatures are now expected to 
distinguish between such models.
 The discussion of  our zero field data will be limited, as in all 
recent
theoretical 
\cite{Graf96,Graf96_2,Fledderjohann95,Norman96,Barash96} and 
experimental \cite{Huxley95,Suderow95,Lussier94,Lussier95} 
work, to two plausible symmetries
for the OP that appear consistent with the observed properties of 
superconductivity in UPt$_3$ : the E$_{1g}$ and E$_{2u}$ 
representations of the order  parameter which both lead  to a 
hybrid gap, with a line node  in the basal plane and nodes along the 
c axis where the gap vanishes respectively  linearly (E$_{1g}$) or 
quadratically (E$_{2u}$) with the polar angle $\theta$.
Up to now, no experimental distinction could be made between the 
two models on the basis of the
thermal conductivity. Since the work of K. Behnia et al. 
\cite{Behnia91}  measurements (Lussier et al. \cite{Lussier94} for 
T$\ge$100mK) emphasizing the appearance of an anisotropy of the 
thermal conductivity below T$_c$ concluded that the gap was of the
E$_{1g}$ type, and further measurements by the same group to 
lower temperatures (T$\ge$50mK) \cite{Lussier95} were 
interpreted to favour E$_{2u}$. We \cite{Huxley95} measured the 
thermal conductivity down to 35 mK, and showed that although 
some features of our data, like the power law behaviours,  
seemed to favour the E$_{2u}$ model, others did not, like the temperature 
dependence of the anisotropy.
 In fact it was shown 
\cite{Graf96,Graf96_2,Fledderjohann95,Norman96,Barash96} on the 
basis of the data of Lussier et al. \cite{Lussier95} that the existing 
experimental data could not differentiate between the two models.
 An
important point is that for the lowest temperatures, which should 
be best suited to 
distinguish between the two gap structures, theory predicts that the 
pair breaking effect of impurities and defects produces a band of 
impurity bound states which restores a behaviour of the thermal 
conductivity comparable to that of the normal state. This regime 
was never observed up to now, as the measurements were not 
extended to low enough values of T/T$_c$.
 Our data down to 16 mK show the onset of this type of
behaviour, which is accompanied by drastic changes on the 
anisotropy of $\kappa$ below 30 mK.
 In addition, the
experimental resolution has been considerably improved, thus our 
data give a much better basis for a more detailed discussion of the 
thermal conductivity. We discuss extensively
our zero field data, and also present  new measurements under 
magnetic fields.
Measurements under field in the mixed state give an additional 
 means of probing the gap structure 
\cite{Behnia91,Behnia92,Behnia92_b,Yin93,Yin94}. Indeed, the 
possibility of a hybrid gap such as that in the E$_{1g}$ and E$_{2u}$ 
models
was proposed by Yin and Maki \cite{Yin93} on the basis of the 
mixed state thermal
conductivity data of Behnia et al. \cite{Behnia91,Behnia92}.  Our 
data are more systematic and show new features of $\kappa$ in the 
mixed state, arising both at very low temperatures and near the 
tetracritical point, where the three superconducting phases of 
UPt$_3$ coalesce.

\section {Experimental}

The crystals used for the measurements were cut at adjacent sites 
of the  same mother crystal and followed exactly the same heat 
treatment. Their critical temperature T$_c$ is the same, their
 specific heat shows \cite{Huxley95} clearly the 
double superconducting transition and their residual  resistivities 
are among the
lowest ever reported ($\rho_0=0.54 \mu\Omega cm$ for the 
current j//b, and $\rho_0=0.17 \mu\Omega cm$ for j//c 
\cite{Huxley95}). The samples were heated by a PtW alloy strain 
gauge heater, and the temperature  was measured by two matched 
Matshushita 68$\Omega$ carbon thermometers. The accuracy of the  
measurements has been significantly improved. We could achieve a 
1\% resolution at 500 mK, 3\% at 50 mK and 10\% at 16mK for the 
measured $\it{anisotropy}$ $\kappa_c/\kappa_b$. By comparison, 
the authors of Ref.\cite{Lussier95} give error bars of 1\% at 150 mK 
and 5\% at 50 mK on $\kappa$, whereas for the anisotropy 
$\kappa_c/\kappa_b$, the error bars are estimated to be 15\% 
above 50 mK\cite{Norman96}. We improved our experimental method 
(described in Ref.\cite{Huxley95}) in the following points: (i) 
better electronics, and extensive time averaging to 
 reduce 
noise. As a consequence, a continous run from 16mK to 1K fully 
computer controlled took 3 days of measuring time. (ii) The long 
time averaging was possible  due to the very high temperature 
stability of our dilution refrigerator, and its low   
temperature  minimum : 5 mK. (iii) In this dilution refrigerator we could 
calibrate 
our thermometers during the same experiment against a Ge 
thermometer, a
paramagnetic CMN salt and a $^3$He vapor  pressure thermometer.
 The lowest measuring temperature for our
samples was limited to 15 mK by their self  heating due to the U 
radioactivity.
 The problem of
the magnetoresistance of the thermometers (the improved 
sensitivity is payed for by important
changes of the calibration under magnetic field) was solved by 
making a new calibration against a standard Ge thermometer in the 
zero  field compensated region of the magnet at every field.
 The
experimental error bars become more important when increasing 
the magnetic field, and will be shown in the figures when 
necessary. In high magnetic fields (3T), and for j//c, the thermal 
conductivity of UPt$_3$ is extremely large : 3 
orders of magnitude larger than at 16mK and zero field, due to the 
low value of the residual resistivity $\rho_0$ for j//c 
($\rho_0=$0.17 $\mu\Omega$cm).
 This, and the poor geometric factor of our sample for j//c
(roughly 2 mm length , compared with 6 mm for the sample 
measured for j//b) conspired to make the measurements at 3 T and 
very low temperatures extremely difficult. Therefore, at 3 T and for 
j//c we only measured above 100mK. Taking another sample with 
better geometry would have solved this problem, but our interest 
was focused on the superconducting phase of UPt$_3$. More than 
60 $\kappa$(T) runs under  different magnetic fields were done in 
order to scan the B-T phase diagramm. All magnetic fields where 
applied along the same axis as the heat current (B//j//b and 
B//j//c).

\section{Normal Phase}

The normal phase behaviour of the thermal conductivity of UPt$_3$ 
shows some features
important for the understanding of the superconducting phase. In 
Figure 1 we have plotted $\kappa(T)$ for j//b and j//c, at 0 T and 
at 3 T where the normal phase is recovered. The dominant 
scattering mechanisms of the electrons are the same as in the 
electrical conductivity\ :
 (i) elastic scattering of electrons by impurities and (ii) inelastic 
scattering between electrons. The first scattering mechanism (i) 
leads to a linear temperature dependence of the thermal 
conductivity and is related to the temperature independent part of 
the resistivity by the Wiedemann-Franz law : 
$\kappa/T=L_0/\rho_0$ ($L_0=2.44$ $10^{-8} W/K^2cm$). As seen 
in Fig.\ 1 this mechanism dominates at very low temperatures 
T$\le$100mK, below which $\kappa/T$ only varies by 10\% as 
T$\rightarrow$0K.  The ratio of the  thermal 
conductivities extrapolated down to T=0K along the b and the c axis is 3. This 
ratio is usually said  
\cite{Lussier94,Lussier95,Graf96,Graf96_2,Fledderjohann95,Norman96}
 to be related to the anisotropy of the Fermi velocities, rather
than to the anisotropy of the scattering rate $\tau$, which is 
assumed to be isotropic.
 Note that this point is not  experimentally demonstrated.
 Nevertheless we will also take $\tau$ to be
isotropic, in order to compare our data with theory.
The second mechanism (ii) leads to the well known Fermi liquid 
T$^2$  behaviour of the electrical
resisitivity  and is of growing importance with increasing 
temperature. Deviations from the simple Wiedemann-Franz law are 
then expected : one expects again 
$\kappa/T=L/(\rho_0+AT^2$) but now with a temperature 
dependent Lorentz number L(T)$\le L_0$.

 In
order to verify this point, we also measured the electrical resistivity 
of our samples at 3 T. In high purity samples, $\omega_c\tau$ 
($\omega_c$ being the cyclotron frequency) becomes
larger than 1 at low temperatures.
This
complicates the temperature dependence of the resistivity at 3 T, so 
that we did not get the expected simple Fermi liquid behaviour 
$\rho=\rho_0+AT^2$. Such effects have been already observed in 
UPt$_3$ \cite{Taillefer88,Kambe96}.
 Nevertheless we can roughly estimate that the Lorentz number 
varies from 0.95 L$_0$  at T=0K to 0.8 
L$_0$ at T=600 mK. We note that the positive magnetoresistance of 
UPt$_3$ ($\rho_0$ varies for j//b (j//c) by roughly 20\% (10\%) 
between 0 and 3 T \cite{Behnia91,Kambe96}) leads one to expect 
that  $\kappa$ decreases with B, as was already observed by 
Behnia et al.\cite{Behnia91} but contrary to what was shown by 
Lussier et al. \cite{Lussier95}, although these authors might have corrected 
qualitatively for the magnetoresistance, assuming the validity of the 
Wiedemann-Franz law in all the measured temperature range.

\section{Superconducting phase}

The thermal conductivity in zero field is shown in Fig.\ 1.
 No significant change is observed at
T$_c$. It is important to realize that up to now quantitative 
comparison with theory in the 
superconducting phase can only be done easily when elastic 
scattering with impurities prevails. As shown in Fig.\ 1, $\kappa$ at 
T$_c$ is roughly half the value of the zero temperature
extrapolation of $\kappa/T$ in the normal state. This means that at 
T$_c$
both mechanisms (i) and (ii) are of equal importance.  When 
entering the superconducting state, $\kappa$ rapidly drops as the 
superconducting gap opens over the Fermi surface. Therefore, the 
inelastic scattering rate in the superconducting state  should 
diminish more rapidly than in the normal state.
 A qualitative treatment  of the
inelastic scattering rate has been proposed by Fledderjohann and 
Hirschfeld \cite{Fledderjohann95}, extrapolating the normal phase 
scattering rate taken to be proportional to $1/(\rho_0 + AT^2)$.
 Norman and Hirschfeld \cite{Norman96} also made an extensive
analysis of Fermi surface effects on $\kappa$, including inelastic 
scattering, and demonstrated the difficulty of fitting $\kappa$ in 
this temperature range.
As suggested in Ref.\cite{Fledderjohann95}, they tried to fit at the same time
 $\kappa(T)$ and the 
anisotropy $\kappa_c/\kappa_b$, shown in Fig.\ 2. This quantity 
has little temperature dependence in the normal phase. As shown 
in the inset of Fig.\ 2, our enhanced resolution permits us to 
observe a kink in 
$\kappa_c/\kappa_b$ against temperature at roughly 550 mK, 
which should correspond to the onset of superconductivity. Near 
T$_c$ (for T$\le$T$_c$),  $\kappa_c/\kappa_b$ is independent of 
temperature, as shown in Ref.\cite{Huxley95}, but contrary to what 
was observed by Lussier et al. \cite{Lussier94,Lussier95}.
 Below
0.8T$_c^-$ (T$_c^-$=480 mK) $\kappa_c/\kappa_b$  starts to 
decrease and seems to tend towards a finite zero temperature 
extrapolation. However below 30 mK (0.07T$_c^-$) a stronger 
dependence on temperature sets in.  Norman and Hirschfeld 
calculated $\kappa_c/\kappa_b$ using many different functions to 
describe the OP within the E$_{1g}$ and E$_{2u}$ irreducible 
representations,
 and one of the curves they calculate (Fig. 8 of Ref.\cite{Norman96})
 fits our measurement of the anisotropy $\kappa_c/\kappa_b$ well,
however their calculated temperature dependence of $\kappa_b$ 
and $\kappa_c$ do not correspond with our data.
 Nevertheless, we note that a temperature independent anisotropy
$\kappa_c/\kappa_b$ near T$_c$ seems to be only reproduced 
within E$_{2u}$ in their calculations. At the lowest temperatures 
(T$\rightarrow$0K) the thermal excitations which drive the heat 
current have $\vec{k}$ vectors corresponding only to the Fermi 
surface regions near the nodes of the gap, and inelastic scattering 
can be ignored, so that the interpretation of $\kappa$ is expected to 
be simpler. We now discuss impurity effects which turn out to be 
important in this limit.

In non conventional superconductors with nodes of the gap, the pair 
breaking effect
of even non magnetic impurities or defects  leads to a finite 
density of states at the Fermi level, due to virtual bound states 
forming a band of width $\gamma$ (see 
Refs. 
\cite{Pethick86,Schmitt86,Monien87,Hirschfeld86,Hirschfeld88,Graf96,Graf96_2,Fledderjohann95,Norman96,Lee93,Borkowski95,Preosti94,Balatsky95}).
 This band is easier to observe 
when the impurities are in the unitarity
scattering limit (scattering phase shift $\delta =\pi /2$), and in 
heavy fermion superconductors, impurity scattering is expected to 
be in this limit   \cite{Pethick86}. Thus, in UPt$_3$,  the 
thermodynamics is expected to be governed by the propeties of this 
band for  T$\le\gamma$.

Before discussing the peculiar properties of this band which leads 
(see
below) to a finite $\kappa/T$ for T$\rightarrow$0K, we focus on 
the power law behaviour expected for intermediate temperatures: 
$\gamma \le T \ll T_c$. In this temperature region, the impurity 
induced band can be neglected, as well as inelastic scattering 
effects.
In the case of resonant scattering of the thermal 
excitations by impurities, simple
power laws are expected, controled by  the gap 
nodes. Our measurements of $\kappa/T$ below 80 mK are shown in 
Fig.\ 3 as a function of T$^2$.
 We can observe  two regimes : (i) between 80 
mK and 30 mK a 
T$^3$ power law behaviour of $\kappa$ for both axis, as expected for
a hybrid gap with resonant scattering 
 and (ii) below 30 mK the 
effects of the impurity induced band of quasiparticles leading 
 to a finite value of $\kappa/T$ for T$\rightarrow$0K.
 Note that the extrapolation
of the data for T$\ge$30mK gives a negative zero temperature
extrapolation for j//c (zero for j//b), which implies that the 
anisotropy $\kappa_c/\kappa_b$ still depends on temperature 
down to 30 mK (see Fig.\ 2).
 In an earlier paper
\cite{Huxley95} we argued that observing the same power laws for 
both axis is in favour of the E$_{2u}$ model, as the quadratic 
angular dependence of the nodes near the c axis leads to a high
density of thermal excitations comparable to the density in the 
basal plane \cite{Fledderjohann95}. Nevertheless, the theoretical 
analysis of the authors of Refs. 
\cite{Graf96,Graf96_2,Norman96,Barash96} demonstrated that this 
observation could also be consistent with an E$_{1g}$ state.
Barash and Svidzinsky \cite{Barash96} examined the expected 
power laws in detail, and they find to leading order in T:

- for j//b (heat current in the basal plane ):
$\kappa_b= T^3 (a_b + b_b ln^2(T/\Delta_0) + c_b ln(T/\Delta_0))$ 
for E$_{1g}$ and E$_{2u}$;

- for j//c (heat current perpendicular to the basal plane):
$\kappa_c= T^3 (a_c + b_c ln^2(T/\Delta_0) + c_c ln(T/\Delta_0))$ 
for E$_{2u}$ and $\kappa_c=a_c T^3$ for E$_{1g}$;

with coefficients $a, b$ and $c$ which depend on the 
parametrization of the gap
near the nodes, and even on contributions of the order parameter 
from regions of the Fermi surface far from the nodes 
\cite{Barash96}.
 The power laws expected on 
$\kappa_c$ for E$_{1g}$ and E$_{2u}$ differ mainly by a 
logarithmic factor ($ 1 + \frac{b_c}{a_c} ln^2(T/\Delta_0) + 
\frac{c_c}{a_c} ln(T/\Delta_0$)).
As seen in Figure 3, both $\kappa_b$ and $\kappa_c$ do not follow 
a pure power law,
but need a negative zero temperature extrapolation to be fitted. The 
formulas given above
(for E$_{2u}$) fit our data well, but with negative
coefficients b$_b$ and b$_c$, which may not have a clear physical 
meaning. Therefore, on
the basis of power law behaviours we are not able to discriminate 
unambiguously between E$_{1g}$ and E$_{2u}$.

The lowest temperature regime (T$\le\gamma$) might still be used 
for the identification of the order parameter. Indeed theoretical 
work has shown \cite{Graf96,Graf96_2,Norman96} that the 
existence of the impurity induced band of quasiparticles does not 
wash out completely the sensitivity of the thermal conductivity to 
the choice of the OP.  We will follow
closely the discussion of Graf et al. \cite{Graf96_2}. They predict for
$T\le\gamma$ : $\kappa_i/T=\alpha_i + \beta_i T^2$ with 
$\alpha_i$ and $\beta_i$ which depend
 on the choice of the OP and of the direction of the heat current :

- For i=b (heat  current in the basal plane)
$\kappa_b$ is the same for E$_{1g}$ and E$_{2u}$, as both predict a 
line node in the basal plane. Moreover, $\alpha_b$ is universal, i.e.\ 
independent of the impurity concentration $n_i$ although it 
depends on the form of the gap near the node, and   $\beta_b = 
\frac{7 \pi^2 k_B^2}{60\gamma^2} \alpha_b$.

- For i=c,
$\alpha_c \propto \gamma$ and $\beta_c=2.5 
\frac{\beta_b}{\alpha_b}\alpha_c$ for E$_{1g}$,
but for E$_{2u}$, $\alpha_c \propto \alpha_b$ is universal and 
$\beta_c=\frac{\beta_b}{\alpha_b}\alpha_c$.

The inset of Fig.\ 3 shows $\kappa/T$ as a function of T$^2$ at the 
lowest temperatures (T$\le$35 mK) together with the fits to the 
predicted formulas. For the thermal conductivity in the basal plane, 
the fit gives $\alpha_b=0.18mW/K^2cm$, of the same order of 
magnitude as
$\alpha_b \approx 1mW/(K^2cm)$ estimated in Ref.\cite{Graf96}. 
It also yields $\beta_b/\alpha_b=4.2\ 10^3/K^2$, therefore $\gamma 
\approx 17mK$. Note that with a less pure sample we should have 
been able to observe the expected law over a larger range of 
temperatures. Assuming that the impurity concentration n$_i$ is 
the same for both measured crystals, and having fixed $\alpha_b$ 
and $\beta_b/\alpha_b$ only $\alpha_c$ can be varied as a free 
parameter in order to reproduce the measured curve for j//c 
\cite{Graf96}. For E$_{2u}$ the best agreement is obtained with 
$\alpha_c=0.2mW/K^2cm$ and for E$_{1g}$ with 
$\alpha_c=0.11mW/K^2cm$.
As shown in Fig.\ 3, the fit to E$_{1g}$ is better, but only
in a small temperature range ($16mK\leq T\leq 25mK$). 
Nevertheless, the assumption that $\gamma$ is the same for both 
measured samples may not be justified. We recall that the 
assumption of an isotropic scattering rate may not hold, and that 
even if both samples come  from the same mother  crystal  and  had  
exactly  the  same  heat treatment,  differences in the impurity (or 
defect) concentration can {\it never} be  excluded. Indeed, although 
the onset of a finite resistivity occurs at the same temperature for both samples, the 
width of the superconducting transition measured 
by resistivity is 17mK for j//b, but only 8mK for j//c (10\%-90\% 
criterion).
 The broadening of the superconducting
transition could be correlated to the sensitivity to stress produced 
in the cutting of
the material despite the subsequent annealing. Note that our 
measurements under magnetic fields show a clear anomaly in 
$\kappa$ at T$_c$. As shown in the inset of Fig.\ 4 this anomaly 
coincides with the onset of a finite resistivity, thus with the lowest part 
of the resistive transition. Therefore, differences in the transition 
width, although they show that the samples are different, do not 
give much information about the defect concentration in the bulk of 
the material.  We note that  in
order to explain our data within   E$_{2u}$, the impurity  
concentration n$_i\propto \gamma^2$ \cite{Graf96} should be 2.5 
times lower  for  j//c  than for j//b. One way of
solving this  problem   is   to  measure less pure samples (in order 
to observe the expected laws in a larger range of temperatures), 
again for both axis and down to the lowest possible temperatures. 
This will also give an experimental test of the prediction of a 
universal value of $\kappa/T(T\rightarrow0K)$ for the heat 
current in the basal plane. Note that the upper critical field 
anisotropy (B$_{c2}$ is slightly Pauli  limited for B//c but not for 
B$\perp$c)  is an important argument in favour of an odd parity 
 OP in UPt$_3$ (see Ref.\cite{Choi91}). It will be 
very difficult to explain B$_{c2}$ within E$_{1g}$.

One important point not understood at present is the discrepancy 
between the value of
$\gamma$ estimated from the normal phase data (with the impurity 
scattering rate $\Gamma_0/T_c \approx 5\ 10^{-2}$ we obtain 
$\gamma \approx 50mK$; see Ref.\cite{Graf96} for the 
formulas), which is 3 times larger than the one found from our low 
temperature fit. Indeed, as pointed out by Norman and Hirschfeld 
\cite{Norman96} the data at higher temperatures ($\gamma \ll T 
\le T_c$) seem to be  difficult to fit with such a low impurity 
scattering rate. This may suggest that the description of the 
impurities and defects in UPt$_3$ only in terms of s-wave 
scattering in the unitarity limit is too simplified for the discussion 
of subtle very low temperature effects. It also stresses the lack of 
experimental knowledge on the nature of the dominant scattering 
centers in this compound.

\section{Mixed Phase}

Figure 4 shows an example for our temperature scans at different 
magnetic fields. The first striking feature is that under
magnetic fields an anomaly appears, which was not visible in zero 
field.
 More precisely, there is a jump
in the derivative of $\kappa$ : d$\kappa$/dT at T$_c$. This is 
predicted to happen already in conventional s-wave 
superconductors, and the interesting point is that, from the 
superconducting side, d$\kappa$/dT is  related to the order 
parameter. At intermediate temperatures (100 
mK$\le$T$\le$T$_c$), $\kappa$ first decreases at low fields as a 
function of B, and then increases up to its value in the normal phase 
(this is seen more clearly in the Fig.\ 8, where we plot $\kappa$ as 
a function of B). At very low temperatures (T$\le$50mK), no 
decrease at low fields is observed and $\kappa$ increases 
continuously with B. Qualitatively speaking, there are 
two competing effects : an enhancement of the thermal conductivity 
with the magnetic field towards the normal phase value, and the 
scattering of the thermal excitations present outside the vortex 
cores by the vortices which may tend to diminish $\kappa$ 
(compared to the zero field value).
 The understanding of the thermal
conductivity of clean superconductors in the mixed state is difficult 
due to the complex interplay between these effects, and new  
theoretical calculations are needed in order to
understand $\kappa$(B). Nevertheless some  conclusions can be 
drawn.

All of these effects strongly depend on the relative direction of the 
magnetic field and
the heat current. Our choice was to measure $\kappa$ with the field 
always parallel to the heat current, along the c axis and in the basal 
plane, in order to make the interpretation easier. Also, the magnetic 
field was always changed in the normal phase (field cooled). We 
checked that only a small irreversibility exists at very low fields 
(B$\le$0.03T).
 
Before discussing the data, we note that we 
recover
the phase diagram of UPt$_3$ when plotting T$_c$ obtained by the 
anomaly of $\kappa$ as a function of the magnetic field (see  Fig.\ 5). The 
B$\rightarrow$C line could also be observed due to anomalies 
shown in Fig.\ 8 on the field dependence of $\kappa$, but no clear 
signature could be found of the A$\rightarrow$B phase transition. 
We will first discuss the thermal conductivity near T$_c$(B), then 
focus on $\kappa(B)$ as T$\rightarrow$0 K, and finally discuss 
$\kappa(B)$ at intermediate temperatures.

In the literature (see Refs.\cite{Behnia91,Behnia92,Yin93,Yin94} for 
the discussion of UPt$_3$), it is generally the jump in 
$d\kappa/dB$ which is  discussed. We have measured 
$d\kappa/dT$ because our resolution for temperature scans was 
much better than for field scans, but the main
conclusions should remain the same. In Fig.\ 6 we show the 
measured $d\kappa/dT$ at T$_c$ as a function of the magnetic 
field, normalized to the normal state values : ($d\kappa_s/dT -
d\kappa_n/dT) / (\kappa_n/T_c(B))$, where the subscript s means 
approaching T$_c$ from the superconducting state, and n from the 
normal state.
 As already mentioned \cite{Yin93,Yin94} and shown by the
measurements of Behnia et al. \cite{Behnia91,Behnia92} in UPt$_3$, 
the slope $d\kappa_s/dB$, and therefore also d$\kappa_s$/dT, 
depends on the direction of current and field, and is related to the 
gap
structure. Therefore, if the topology of the gap changes when 
passing from the high temperature low field A phase to the low 
temperature high field C phase, a jump of $d\kappa/dT$ should be 
observed at the A$\rightarrow$C phase transition. Indeed, as 
shown in Fig.\ 6, we clearly observe this jump for both orientations 
of the field at the A$\rightarrow$C phase transition (0.5 T for B//b 
and 1 T for B//c).
 If no phase transition took place, the overall
behaviour of the measured quantity would be parabolic like, with 
$(d\kappa_s/dT - d\kappa_n/dT) /(\kappa_n/T_c(B)) = 0$ at B 
$=$0 and at $B=B_{c2}(T=0)$, with a maximum at an intermediate 
field. We expect this parabola to depend on the topology of the 
superconducting gap. So it 
might be possible to draw more conclusions
about the gap structure of the A and the C phases, from which little 
is known, by a careful theoretical analysis of these data.

At very low temperatures (T$\leq$50mK) we observe a continuous 
increase (Fig.\ 7) of $\kappa(B)$ up to its normal phase behaviour, 
which we will discuss now. First, we recall the situation in clean s-wave 
superconductors. Within a naive model, one might think that 
in the limit T$\rightarrow$0K the contribution to the thermal 
conductivity coming from the electronic excitations within the 
vortex cores would dominate $\kappa$, as this type of excitation 
dominates the density of states (in s-wave superconductors). One 
would then expect that $\kappa$ scales roughly as 
$\frac{B}{B_{c2}} \kappa_n$, but this turns out not to be correct. 
Indeed, the low group velocity of the excitations within the vortex 
cores leads to a small contribution of
this type of excitations to the thermal conductivity. Vinen et al. 
\cite{Vinen71}
estimate $\kappa_{vortex}$ $_{cores} \approx 10^{-2} 
\frac{B}{B_{c2}} \kappa_n$ for pure Nb. We observe in UPt$_3$ an 
increase of $\kappa$ with the
magnetic field which scales roughly as
 $\frac{B}{B_{c2}} \kappa_n$ for B//j//b and as 
$\frac{1}{3}\frac{B}{B_{c2}} \kappa_n$ for B//j//c at low fields 
(Fig.\ 7). This is a surprising result which is not 
explained with the
existing models.
It might be related to the peculiar magnetic field behaviour of the density of 
states of unconventional superconductors with a line of nodes of 
the gap. Indeed, it has been predicted that in such a case, the 
density of states would be dominated by the contribution of the thermal excitations of 
the superconducting phase outside the vortex cores, leading to 
N(E$_F$) $\propto \sqrt{B}$ instead of N(E$_F$) $\propto B$ 
when the main excitations are those in the vortex cores. Note that this is valid
 at low fields for $B_{c1} < B\ll 
B_{c2}$ \cite{Volovik93,Barash96}. Again, in a naive model, one 
would expect that $\frac{\kappa}{T}$(B) as T$\rightarrow$0K 
follows the density of states, but no signature of the expected 
square root behaviour due to the line of nodes predicted within 
E$_{1g}$ and E$_{2u}$ is found in our data.
 Nevertheless,it is already known that in
the case of clean s-wave superconductors, the thermal conductivity does 
not  in general follow the density of states. We note also that the 
thermal excitations outside the vortex cores are scattered by the 
vortices (see below), and this may considerably influence 
$\frac{\kappa}{T}$(B) also in the limit T$\rightarrow$0K. A rough 
square root behaviour was found in the specific heat C by Ramirez 
et al. \cite{Ramirez95}, although at suspiciously high temperatures (150 
mK; T$_c$/3), and large fields (up to B$_{c2}$/3).
 These measurements are therefore influenced by the field 
dependence of T$_c$ and 
by a huge low
temperature anomaly present in C \cite{Brison94,Schubert92}. The 
origin of this anomaly, which appears 
already at 100 mK in zero field is not understood at present, but it is
for sure not connected with the superconducting state 
\cite{Schubert92}. So the measurements of Ramirez et al. 
\cite{Ramirez95} are clearly not simply related with the field 
dependence of the density of states at T$\rightarrow$0K.
 The advantage of the thermal conductivity is that it seems not to 
be
influenced by this anomaly.
Its origin may therefore be related to localized modes which do not 
carry heat \cite{Huxley95}.
In summary, it is clear that  $\kappa(B)$ as T$\rightarrow$0K has 
a peculiar behaviour in UPt$_3$ not explained by  the 
existing models. One must explain not only  
the linear behaviour of $\kappa$ with B at low fields, but also the 
anisotropy found for the constants of proportionality ($\kappa 
\simeq B/B_{c2}$ for B//j//b and $\kappa \simeq \frac{1}{3} 
B/B_{c2}$ for B//j//c).

As $\kappa(B)$  scales roughly as $\frac{1}{3} 
\frac{B}{B_{c2}}\kappa_n$ for B//j//c, $\kappa(B)$ cannot remain 
linear in B up to B$_{c2}$. Indeed, at roughly 0.8B$_{c2}$, 
$\kappa_c$ strongly deviates from the linear behaviour in order to 
reach the normal state value at B$_{c2}$, whereas $\kappa_b$ is 
roughly linear in the whole field range up to B$_{c2}$.
This behaviour is observed for the extrapolation T$\rightarrow$0K, 
as well as at 100 mK, as
shown in Fig.\ 8. Note that Behnia et al. \cite{Behnia91} already 
observed a large value for d$\kappa$/dB  with B//c when the 
upper critical field is Pauli limited along the c
axis \cite{Choi91}. This may influence $\kappa$(B) near B$_{c2}$.
This measurement, and the ones of Behnia et al. (Refs. 
\cite{Behnia91,Behnia92} in UPt$_3$ and Ref. \cite{Behnia92_b} in 
URu$_2$Si$_2$, another heavy fermion superconductor with a Pauli 
limited upper critical field along one crystallographic axis)
are to our 
knowledge  the only existing thermal conductivity measurements
 in the mixed phase of superconductors with a Pauli 
limited upper critical field. We could not find theoretical predictions 
in this case for $\kappa(B)$ near B$_{c2}$.

We now focus on the magnetic
field behaviour of $\kappa$ between 100 mK and T$_c$, shown in 
Fig.\ 8 and obtained by plotting our temperature scans as a function 
of B. As pointed out above, the scattering of thermal excitations by 
vortices leads to a diminution of $\kappa(B)$ visible at low fields. 
This type of scattering was indeed reported experimentally in s-
wave
superconductors (see e.g. Refs.\cite{Chakalskii78,Lowell70}), and in 
UPt$_3$ by Behnia et al. \cite{Behnia91} and Huxley et al. 
\cite{Huxley95}.
 In a simple model for low fields where vortices can be treated as 
independent
scattering centers,  the additional thermal resistivity 
W=1/$\kappa$ (Fig.\ 9) coming from
scattering of the superconducting thermal excitations by the 
vortices is expected to be 
proportional to the magnetic field (see e.g. Refs.\cite{Chakalskii78,Lowell70,Clearly70}).
 In s-wave
superconductors, the law W = W(0)(1+$\frac{\sigma 
l_0}{\Phi_0}B$) ($l_0$ is the mean free path in the superconducting 
phase, $\Phi_0$ the flux quantum, and $\sigma$ the effective 
scattering width of the vortices)  is observed at low fields and in a 
large range of temperatures.  Our data  
do not show such a behaviour.
 To understand this, the field dependence of $\kappa$ for
T$\rightarrow$0K has first to be explained.
 The fact that we do
not observe the classical behaviour may indeed be related to the 
unconventional nature of superconductivity in UPt$_3$.

\section{Conclusion}

We have shown that the thermal conductivity shows a rich 
behaviour in the different superconducting phases of UPt$_3$.
Experimentally, we could reach low enough temperatures to enter 
the regime where the thermal
conductivity is governed by impurity bound states.
 It is unclear if it is now  possible to differentiate between 
E$_{1g}$ or E$_{2u}$ on the basis of this data, we can however 
parametrize the gap nodes for both models.
	It appears that measurements on samples of different purity 
down to the same lowest temperatures would be very useful. First, they could test 
the validity of the theoretical models, predicting for E$_{1g}$ as 
well as for E$_{2u}$ a universal value of $\kappa/T$ for 
T$\rightarrow$0K in the basal plane. Second, they might distinguish 
between  E$_{1g}$ and E$_{2u}$ : theory predicts for the heat 
current along the c axis a universal value of $\kappa/T$ for 
T$\rightarrow$0K in the case of E$_{2u}$, but a sample dependent 
value for E$_{1g}$. Although our very low temperature data are slightly better
explained within E$_{1g}$, we remind the reader that E$_{1g}$  has 
difficulties to  explain other measurements on UPt$_3$ such as the 
upper critical field \cite{Choi91} or the Knight shift \cite{Tou96} 
which both point towards an odd parity OP.

We do not observe any coupling between the low temperature 
specific heat
anomaly and the thermal conductivity at very low temperatures. 
This points to a localized nature of the modes that lead to this specific heat
anomaly. As indicated by Fomin and Flouquet 
\cite{Fomin96} it could be possible that the real magnetic ordering 
in UPt$_3$ arises only at much lower temperatures ($T\approx$ 20 
mK) than indicated by the appearance of a neutron diffraction 
signal (at $T\approx$ 5K). This signal could indeed be related to the 
onset of short range order as a consequence of large magnetic 
fluctuations, rather than to a real long range magnetic order 
\cite{Fomin96}. But it can not be excluded that a weak 
interplay between these local moments and the heavy electrons 
may influence the temperature dependence of $\kappa$ and 
complicate a quantitative fit.

 We have also presented measurements of the overall temperature and 
field behaviour
of the thermal conductivity in the mixed state. The low field low
temperature result shows the absence of the expected square root 
behaviour of the thermal conductivity, for both measured axis. We 
have also seen that the particular
field dependence of $\kappa(B)$ near B$_{c2}$ is very different 
along the b or c axis axis ($d\kappa_c/dB \gg d\kappa_b/dB$), 
which may be related to the Pauli 
limitation of the upper critical field along the c axis. Clearly, both 
points invite more theoretical calculations and might be deeply 
connected to the OP of UPt$_3$. In the intermediate field regime, 
both the vortex core contribution to $\kappa$ and quasiparticle-vortex 
scattering are important, and the data are difficult to 
interpret without further theoretical work.
	Near T$_c$, the
derivative of the thermal conductivity, directly related to the order 
parameter, shows a clear jump when crossing the A$\rightarrow$C 
phase boundary. This is a clear signature of the change of  
symmetry of the OP at this phase transition.

Acknowledgements.
We would like to thank very useful discussions with P. W\"{o}lfle, P. 
J. Hirschfeld, 
A.I. Buzdin and V.P. Mineev.

\begin{figure}

\caption{The thermal conductivity of UPt$_3$ below 1 K for the 
heat
current along the b and along the c axis. The zero field data are 
shown together with the normal phase data, measured at 3 T.} 
\end{figure}

\begin{figure}
\caption{The anisotropy $\kappa_c/\kappa_b$ normalized to 1 at 
T$_c^-=480mK$ is shown as a function of the reduced temperature 
T/T$_c^-$. The inset shows a kink of $\kappa_c/\kappa_b$ 
at T$_c^+$=550 mK.
Note the low error bars compared with earlier work
(Refs. \protect\cite{Behnia91,Huxley95,Lussier94,Lussier95}).} 
\end{figure}

\begin{figure}
\caption{The very low temperature thermal conductivity of 
UPt$_3$ (between 70mK and 16mK) as a function of T$^2$, with the 
fits based on the predictions of Graf et al. 
\protect\cite{Graf96}. The inset shows $\kappa/T$ between 34mK 
and 16mK. Below 30 mK, the properties of the impurity induced 
band of quasiparticles are observed (see text).} 
\end{figure}

\begin{figure}
\caption{Some of our temperature scans at constant magnetic fields 
are shown in this figure. The inset shows the observed anomaly at 
T$_c$, compared to the resistive transition at 1.56 T.}
\end{figure}

\begin{figure}
\caption{Phase diagramm of UPt$_3$ deduced from the thermal 
conductivity measurements. The A$\rightarrow$N and 
C$\rightarrow$N transitions are traced by the anomaly on the 
temperature scans which appear at T$_c$ under magnetic field. The 
B$\rightarrow$C phase
transition is seen on the field dependence of $\kappa$ (see Fig.\ 8). 
No clear sign of the A$\rightarrow$B transition was seen in our 
measurements. Lines are guides to the eye.}
\end{figure}

\begin{figure}
\caption{The slopes obtained with our temperature scans under 
magnetic fields. Lines are guides
to the eye : continuous lines in the A phase, and dashed dotted lines 
in the C phase. The error bars, which become more important at 
high fields due to the broadening of the transition are also shown.} 
\end{figure}

\begin{figure}
\caption{The figure shows $\kappa/T(B)$ for T$\rightarrow$0K at low fields.
Note that surprisingly, $\kappa_b$ and $\kappa_c$ show a linear behaviour (see text).
At very low fields, close 
to B$_{c1}$ a curvature of $\kappa(B)$ is observed, but in this 
region a non negligible irreversibility exist.} \end{figure}

\begin{figure}
\caption{The magnetic field dependence of $\kappa/T(B)$ at some 
representative temperatures together with the zero temperature 
extrapolation of $\kappa/T(B)$ (open circles). Note that, while 
$\kappa_b/T$ is roughly linear with the magnetic field, 
$\kappa_c/T$ shows a strong upturn near B$_{c2}$ up to the 
normal phase value.} \end{figure}

\begin{figure}
\caption{The magnetic field dependence of the thermal resistivity 
$W(B)=1/\kappa(B)$ at low fields normalized to the value at zero 
field W$_0$. Lines are guides to the eye. Note that our 
measurements, more extensive than in 
Refs.\protect\cite{Behnia91,Huxley95}, show that $W$ is not linear 
in B at low fields (see text).} \end{figure}
\end{document}